\documentclass{IEEEtran}
\IEEEoverridecommandlockouts
\usepackage{cite}
\usepackage{algorithmic}
\usepackage{xcolor}
\usepackage{enumerate}
\usepackage{amsfonts,amsmath,amssymb,amsfonts}
\usepackage{amsthm}
\usepackage{algorithmic}
\usepackage{optidef}
\usepackage{algorithm}
\usepackage{graphicx}
\usepackage{textcomp}
\usepackage{subfigure}
\usepackage[squaren]{SIunits}
\usepackage[section]{placeins}
\def\BibTeX{{\rm B\kern-.05em{\sc i\kern-.025em b}\kern-.08em
    T\kern-.1667em\lower.7ex\hbox{E}\kern-.125emX}}
\begin{document}
\newtheorem{lemma}{\textbf{Lemma}}
\bibliographystyle{IEEEtran}
\title{Meta Federated Reinforcement Learning for Distributed Resource Allocation}
\author{Zelin~Ji,~\IEEEmembership{Graduate Student Member,~IEEE}, Zhijin~Qin,~\IEEEmembership{Senior Member,~IEEE}, and Xiaoming~Tao
\thanks{Part of this work was presented at the IEEE International Conference on Communications 2022~\cite{ji2022federated}.}
\thanks{Zelin~Ji is with School of Electronic Engineering and Computer Science, Queen Mary University of London, London E1 4NS, U.K. (email: z.ji@qmul.ac.uk).}
\thanks{Zhijin~Qin and Xiaoming~Tao are with Department of Electronic Engineering, Tsinghua University, Beijing, China. (email: qinzhijin@tsinghua.edu.cn; taoxm@tsinghua.edu.cn).
}
}
\maketitle
\begin{abstract}
In cellular networks, resource allocation is usually performed in a centralized way, which brings huge computation complexity to the base station (BS) and high transmission overhead. This paper explores a distributed resource allocation method that aims to maximize energy efficiency (EE) while ensuring the quality of service (QoS) for users. Specifically, in order to address wireless channel conditions, we propose a robust meta federated reinforcement learning (\textit{MFRL}) framework that allows local users to optimize transmit power and assign channels using locally trained neural network models, so as to offload computational burden from the cloud server to the local users, reducing transmission overhead associated with local channel state information. The BS performs the meta learning procedure to initialize a general global model, enabling rapid adaptation to different environments with improved EE performance. The federated learning technique, based on decentralized reinforcement learning, promotes collaboration and mutual benefits among users. Analysis and numerical results demonstrate that the proposed \textit{MFRL} framework accelerates the reinforcement learning process, decreases transmission overhead, and offloads computation, while outperforming the conventional decentralized reinforcement learning algorithm in terms of convergence speed and EE performance across various scenarios.
\end{abstract}


\section{Introduction}
The inexorable progression of wireless networks is the trend. The 3rd Generation Partnership Project (3GPP) has standardized the access technique and physical channel model for the fifth-generation new radio (5G NR) network, which enables dynamic switching of user equipment (UE) between resource blocks (RBs) possessing varying bandwidths and supports multiple subcarrier spacing~\cite{3gpp_38.211, 3gpp_38.901}. Building upon the foundation established by 5G, the sixth generation (6G) and beyond networks aspire to provide the enhanced and augmented services of 5G NR, while transitioning toward decentralized, fully autonomous, and remarkably flexible user-centric systems~\cite{6G_resource}. These emerging techniques impose more stringent requirements on decentralized resource allocation methods, emphasizing the significance of optimizing RB assignments to enhance the overall quality of service (QoS) within the systems.

Nevertheless, the fast variations and rapid fluctuations in channel conditions render conventional resource allocation approaches reliant on perfect channel state information (CSI) impractical~\cite{7913583}. The inherent non-convexity of the resource allocation problem resulting from discrete resource block association necessitates computationally demanding solutions. Furthermore, the coupled variables further exacerbate the complexity of the problem. Traditionally, resource allocation problems have been addressed through matching algorithms executed at the central base station (BS), resulting in substantial computational burdens on the cloud server. All of the aforementioned challenges require a brand-new optimization tool capable of effectively operating in unstable wireless environments.


Machine learning (ML) methods, especially deep learning (DL) approaches, have become promising tools to address mathematically intractable and high-computational problems. However, artificial neural networks (NNs) require massive amounts of training data, even for a simple binary classification task. Moreover, the overfitting issue makes artificial NNs hard to adapt and generalize when facing new environments, hence requiring additional data to retrain the models and affecting the training data efficiency. Particularly, the fast channel variations and the flexible network structure in 5G beyond network services restrict the application of conventional ML algorithms. 

To enable fast and flexible learning, meta learning has been proposed to enable the model to adapt to new tasks with faster convergence speed by taking the input of experience from different training tasks~\cite{learn_to_learn, fast_RL, MAML2017}. For instance, model-agnostic meta-learning (MAML)~\cite{MAML2017} is a meta-learning technique that can integrate prior experience and knowledge from the new environment, empowering the models with the ability to generalization and fast adaptation to new tasks. Another way to improve data efficiency is to enable experience sharing among models, which is known as federated learning. By the periodic local model averaging at the cloud BS, federated learning enables the local users, to collectively train a global model using their raw data while keeping these data locally stored on the mobile devices~\cite{8994206}. In this paper, we focus on meta learning enabled federated reinforcement learning, to improve the performance of the reinforcement learning algorithm for resource allocation tasks in wireless communications.

Through the implementation of periodic local model averaging at the cloud-based base station (BS), federated learning facilitates collaborative training of a global model by enabling local users to utilize their respective raw data, which remains stored locally on their mobile devices~\cite{8994206}. This paper investigates the application of meta learning within the context of federated reinforcement learning, with the aim of enhancing the performance of the reinforcement learning algorithm in resource allocation tasks within wireless communication systems.

\subsection{Related work}

\subsubsection{Energy-Efficient Resource Allocation for Cellular Networks}

Presently, most cellular user equipment (UE) operates on battery power, and the use of rate maximization-oriented algorithms~\cite{1561930} may result in unnecessary energy consumption, which is unfavorable for the advancement of massive capacity and connectivity in 5G and beyond communications.

Existing literature on energy-efficient resource allocation primarily focuses on optimizing transmit power and channel assignment~\cite{7579565, 4205089}. Robat Mili \emph{et al.}~\cite{7579565} concentrate on maximizing energy efficiency (EE) for device-to-device communications. While numerous studies have investigated resource allocation in wireless communication systems, most of them rely on centralized approaches, which are considered as complex and not easily scalable\cite{4205089}. In such centralized approaches, the central entity needs to obtain global channel state information (CSI) to assign channels to UEs, leading to significant communication overhead and latency. Consequently, distributed low-complexity algorithms are preferable over centralized ones.

Game theory has been adopted for decentralized resource allocation~\cite{4205089, 9280358, 8103020}. However, these approaches typically assume a static radio environment and require multiple iterations for UEs to converge to the Nash Equilibrium (NE) point. In the practical environment, the performance of game theory based algorithms is impacted by the rapid fluctuations in the wireless channel. Yang \emph{et al.} \cite{9280358} and Dominic \emph{et al.} \cite{8103020} integrate the game theory and stochastic learning algorithm (SLA) to enable local users to learn from past experience and adapt to channel variations. Nevertheless, game theory based algorithms do not fully explore the advantages of collaboration and communication among users, potentially affecting system-level performance.


%

%
\subsubsection{Decentralized Reinforcement Algorithms in Wireless Communications}
A promising solution to address concerns regarding complexity and signaling cost concerns involves establishing a decentralized framework for resource allocation and extending the intelligent algorithms to encompass cooperative large-scale networks. The adoption of multi-agent reinforcement learning (MARL) algorithm presents an opportunity to tackle the challenges associated with complexity and enhance the intelligence of local UEs. MARL algorithms rely solely on real-time local information and observations, thereby significantly reducing communication overhead and latency. Mathematically, MARL can be formulated as a Markov decision process (MDP), where training agents observe the current state of the environment at each step and determine an action based on the current policy. Agents receive corresponding rewards that evaluate the immediate impact of the chosen state-action pair. The policy updates are based on the received rewards and the specific state-action pair, and the environment transitions to a new state subsequently. The application of MARL approaches in wireless communications has been extensive~\cite{liang2019spectrum,9026965, RIS_resource}. Wang \emph{et al.}~\cite{9026965} have demonstrated that such a decentralized optimization approach can achieve near-optimal performance. However, local user equipment (UE) cannot directly access global environmental states, and UEs are unaware of the policies adopted by other UEs. Consequently, there is a possibility that UEs may select channels already occupied by other UEs, leading to transmission failures in the orthogonal frequency-division multiple access (OFDMA) based schemes.

\subsubsection{Reinforcement Algorithm for Jointly Resource Optimization}
It is noted that the resource block association problem is a discrete optimization problem, which is usually solved by value-based methods, e.g., Q-learning, SARSA, and Deep Q-learning. Meanwhile, the transmit power is the continuous variable, and only policy-based algorithm can deal with the continuous optimization. Hence, how to jointly optimize the transmit power and channel assignment becomes a challenge. In some work, the transmit power is approximated to discrete power levels, and the user can only transmit by these presetting power levels~\cite{ji2022federated, 8633948}. However, discrete transmit power with large intervals means performance reduction.On the other hand, the complexity could be very high if the number of power levels is significant. To address these concerns, Yuan \emph{et al.} \cite{9495238} proposed a framework with a combination of value-based network and policy-based network. Similarly, Hehe \emph{et al.} \cite{he2021noma} also proposed a combination framework with different components to address the discrete user association problem and continuous power allocation problem. However, in such works the different networks are trained simultaneously, which leads to an unstable framework and makes the NNs hard to be trained and converge.

\subsection{Motivations and Contributions}
\subsubsection{Federated Reinforcement Learning}

The primary obstacle faced by MARL algorithms is the instability and unpredictability of actions taken by other user equipment (UEs), resulting in an unstable environment that affects the convergence performance of MARL~\cite{foerster2017stabilising}. Consequently, a partially collaborative MARL structure with communication among UEs becomes necessary. In this structure, each agent can share its reward, RL model parameters, action, and state with other agents. Various collaborative RL algorithms may employ different information-sharing strategies. For instance, some collaborative MARL algorithms require agents to share their state and action information, while others necessitate the sharing of rewards. The training complexity and performance of a collaborative MARL algorithm are influenced by the data size that each agent needs to share. This issue becomes severer when combining neural networks (NN) with reinforcement learning. In a traditional centralized reinforcement algorithm, e.g., deep Q-network (DQN), the environment's interactive experiences and transitions are stored in the replay memory and utilized to train the DQN model. However, in multi-agent DQN, local observations fail to represent the global environment state, significantly diminishing the effectiveness of the replay memory. Although some solutions have been proposed to enable replay memory for MARL, these approaches lack scalability and fail to strike a suitable balance between signaling costs and performance.

To address the issue of non-stationarity, it is necessary to ensure the sharing of essential information among UEs, which can be facilitated by federated learning~\cite{9530714}. Federated learning has demonstrated successful applications in tasks such as next-word prediction~\cite{hard2019federated} and system-level design~\cite{bonawitz2019federated}. Specifically, federated reinforcement learning (FRL) enables UEs to individually explore the environment while collectively training a global model to benefit from each other's experiences. In comparison to MARL approaches, the FRL method enables UEs to exchange their experiences, thereby enhancing convergence performance~\cite{45895}. This concept has inspired the work of Zhang \emph{et al.}\cite{9348485} in improving WiFi multiple access performance and Zhong \emph{et al.}\cite{zhong2021mobile} in optimizing the placement of reconfigurable intelligent surfaces through the application of FRL.

\subsubsection{Meta Reinforcement Technique for Fast Adaptation and Robustness}
Another main challenge of the reinforcement learning algorithm is the demand for massive amounts of training data. Since the training data can only be acquired by interacting with the environment, the agent usually needs a long-term learning process until it can learn from a good policy. Moreover, using such a large amount of data to train an agent also may lead to overfitting and restrict the scalability of the trained model. In the scope of the wireless environment, the fast fading channels and unstable user distributions also put forward higher requirements on robustness and generalization ability. Particularly, the previous resource allocation algorithms usually set a fixed number of users, which makes the algorithm lack scalability to various wireless environments in practical implementation. 

Meta learning is designed to optimize the model parameters using less training data, such that a few gradient steps will produce a rapid adaptation performance on new tasks. During the meta learning training process, the model takes a little training data from different training tasks to initialize a general model, which reduces the model training steps significantly. The meta learning can be implemented in different ways. Wang \emph{et al.}~\cite{learn_to_learn} and Duan \emph{et al.}~\cite{fast_RL} have applied recurrent NN and the long short-term memory to integrate the previous experience into a hidden layer, and NNs have been adopted to learn the previous policy. Finn \emph{et al.}~\cite{MAML2017} have leveraged the previous trajectories to update the NNs, and further extended the meta learning to reinforcement learning. In this paper, we consider the meta learning for initializing the NNs for MARL. In the scope of wireless communications, Yuan \emph{et al.}~\cite{9495238} have adopted the meta reinforcement learning for different user distributions and confirm that the meta reinforcement leaning is a better initialization approach and can achieve better performance in new wireless environments.

Another challenge caused by federated learning is the heterogeneity in systems and the non-identical data distributions in RL may slow down or even diverge the convergence of the local model. Inspired by the meta learning, Fallah \emph{et al.}~\cite{NEURIPS2020_24389bfe} have developed a combined model, in which the global training stage of the federated learning can be considered as the initialization of the model for meta learning, and the personalized federated learning stage can be seen as the adaptation stage for meta learning. Due to the similar mathematical expression, we can combine federated learning and meta learning naturally, so that training and adapting the models from statistically heterogeneous local RL replay memories. The aforementioned studies serve as valuable inspiration for us to explore the application of meta learning and FRL in addressing the challenges of channel assignment and power optimization. By leveraging these techniques, we aim to distribute the computational load to local user equipment (UEs), reduce transmission overhead, and foster collaboration among UEs.

This paper introduces a novel framework that combines meta learning and FRL for distributed solutions to the channel assignment and power optimization problem. To the best of our knowledge, this is the first endeavor to integrate meta learning and FRL in the context of resource allocation in wireless communications. The contributions of this paper are summarized as follows:
\begin{enumerate}[1)]


\item {A meta federated reinforcement learning framework, named \textit{MFRL}, is proposed to jointly optimize the channel assignment and transmit power. The optimization is performed distributed at local UEs to lower the computational cost at the BS and the transmission overhead.}

\item {To improve the robustness of the proposed algorithm, we leverage the meta learning to initialize a general model, which can achieve fast adaptation to new resource allocation tasks and guarantee the robustness of the proposed \textit{MFRL} framework.}

\item {To address the joint optimization of the discrete and continuous variables, we redesign the action space for the RL algorithm and design the corresponding proximal policy optimization (PPO) network to optimize the real-time resource allocation for each UE.}

\item {To explore the collaboration among cellular users, we propose a global reward regarding the sum EE and the successful allocation times for all UEs and apply the \textit{MFRL} framework for enabling experience sharing among UEs.}
\end{enumerate}

The remainder of the paper is organized as follows. In Section~\uppercase\expandafter{\romannumeral2}, the system model is presented and an EE maximization problem is formulated. The meta federated reinforcement learning algorithm is presented in Section~\uppercase\expandafter{\romannumeral3}. The proposed \textit{MFRL} framework is illustrated in Section~\uppercase\expandafter{\romannumeral4}. The numerical results are illustrated in Section~\uppercase\expandafter{\romannumeral5}. The conclusion is drawn in Section~\uppercase\expandafter{\romannumeral6}.

\section{System Model}

In this paper, we assume that the set of UEs is denoted as ${\cal UE} = \left \{UE_1, \dots, UE_I \right\}$, where $I$ is the total number of UEs. For $UE_i$, the binary channel assignment vector is given by $\boldsymbol {\rho}_i = \left [\rho_{i,1}, \dots,\rho_{i,n}, \dots, \rho_{i,N} \right], i \in I, n\in N$, where $N$ is the number of subchannels. The channel assignment parameter $\rho_{i,n}=1$ indicates that the $n$-th subchannel is allocated to $UE_i$, otherwise $\rho_{i,n} = 0$. Each UE can only accesses one channel, i.e., $\sum \nolimits ^N_{n=1} \rho_{i,n} = 1, \forall i \in I$. Meanwhile, we consider a system with OFDMA, which means a channel can be accessed by at most one UE within a cluster, i.e., $\sum \nolimits^I_{i=1} \rho_{i,n} \in \{0,1\}, \forall n \in N$. In the case of each user equipment (UE), successful transmission with the base station (BS) is achieved when the UE accesses a specific subchannel without any other UEs within the same cluster accessing the same subchannel. Consequently, if each UE is allocated a channel that does not conflict with other UEs within the cluster, this allocation is considered a successful channel assignment.

The pathloss of a common urban scenario with no line of sight link between $UE_i$ and the BS can be denoted by~\cite{3gpp_38.901}
\begin{equation}
PL_{i,n} = 32.4 + 20 \log_{10}\left (f_n \right) + 30 \log_{10} \left (d_{i,n}\right) (\text{dB}),
\label{eq1}
\end{equation}
where $d_{i,n}$ represents the 3D distance between $UE_i$ and the BS, $f_n$ represents the carrier frequency for $n$-th subchannel. Considering the small-scale fading, the overall channel gain can be thereby denoted by
\begin{equation}
    h_{i,n} = \frac{1}{10^{(PL_{i,n}/10)}} \psi m_n,
\label{eq2}    
\end{equation}
where $\psi$ is the log-normally distributed shadowing parameter. According to the aforementioned pathloss model, there is no line of sight between UEs and the BS, and $m_n$ represents the Rayleigh fading power component of the $n$-th subchannel. Hence, the corresponding signal-to-noise ratio (SNR) between the BS and $UE_i$ transmitting over the $n$-th subchannel is represented as 
\begin{equation}
    \gamma_{i,n} = \frac{\rho_{i,n}h_{i,n}p_{i}}{N_n},
\label{eq3}
\end{equation}
where $N_n = W_n\sigma_n^2$ represents the Gaussian noise power on the $n$-th subchannel. The uplink EE for a successful channel assignment of $UE_i$ is given by 
\begin{equation}
u_{i,n} =
\begin{cases}
\frac{BW_n}{p_i} \log_2\left (1+\gamma_{i,n}\right), &\text{if  } \sum \nolimits ^N_{n=1} \rho_{i,n} = 1;\\
0,&\text{else}.
\end{cases}
\label{eq4}
\end{equation}
where $BW_n = k\times b_n$ is the bandwidth of the $n$-th subchannel, $k$ represents the number of subcarriers in each subchannel, and $b_n$ denotes the subcarriers spacing for $n$-th subchannel. Meanwhile, for the unsuccessful assignment, i.e., the UE cannot access any subchannel, the uplink rate is set to $0$ as it is unacceptable for the OFDMA system.



The problem is formulated as
\begin{maxi!}|l|
{\{\boldsymbol \rho, \boldsymbol p\}}{\sum^I_{i=0} \sum^N_{n=0} u_{i,n}}
{\label{eq5}}{(\textbf{P0})}
\addConstraint{ p_i \leq p_{max}, \forall i \in I \label{objective:c1} }
\addConstraint{\gamma_{i,n} > \gamma_{min}, \forall i \in I \label{objective:c2} }
\addConstraint{\sum ^N_{n=1} \rho_{i,n} = 1
, \forall i \in I \label{objective:c3} }
\addConstraint{\sum \nolimits ^I_{i=1} \rho_{i,n} \in \{0, 1\}, \forall n \in N. \label{objective:c4} }
\end{maxi!}
where $\boldsymbol p = \left \{p_1, \dots, p_I \right\}$ denotes the transmit power vector of UEs, $\gamma_{min}$ represents the minimum SNR requirement to guarantee the QoS for UEs. Constraint (\ref{objective:c3}) and (\ref{objective:c4}) make the EE maximization problem a non-convex optimization problem and cannot be solved by mathematical convex optimization tools. In the literature, channel allocation problems are usually formed as linear sum assignment programming (LSAP) problems. To solve this problem, local CSI or the UE related information, e.g., location and velocity should be uploaded to the BS, then the centralized Hungarian algorithm~\cite{3800020109} can be invoked to solve the problem with computational complexity $O\left (I^3\right)$. The computational complexity grows exponentially with the number of UEs, and the mobility of UEs causes the variable CSI, which means the high-complexity algorithm needs to be executed frequently, leading to high transmission overhead and high computational pressure to the BS. Moreover, due to the transmission latency, the current optimized resource allocation scheme by the BS may not be optimal for UEs anymore, and a distributed and low complexity resource allocation approach on the UE side is more than desired.

\textcolor[rgb]{0.00,0.00,0.00}{According to the constraint (\ref{objective:c3}) and (\ref{objective:c4}), each UE can only access one subchannel, and it is clear that the subchannel assignment is a discrete optimization problem. As aforementioned concerns in Section~\uppercase\expandafter{\romannumeral1}, it is hard to train different types of neural networks simultaneously. In another way, the discrete assignment problem can be described by different probabilities to choose different subchannels, and then one-dimensional discrete choice can be mapped to high-dimensional probability distributions. Overall, the joint optimization problem can be solved by a simple policy-based framework with a specific output design.}

\section{Proposed Meta Federated Reinforcement Learning for Resource Allocation}

\begin{figure}[t]
\centering
\includegraphics[width=\columnwidth]{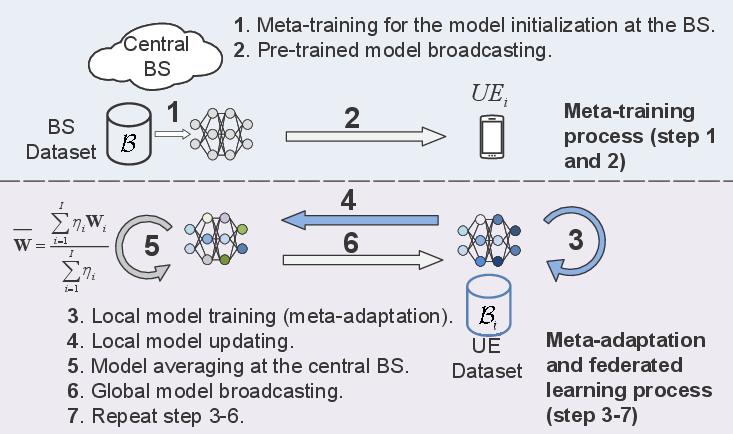}
\centering
\centering
\caption{The proposed \textit{MFRL} framework. The local models are uploaded and averaged periodically.}
\label{FL_framework}
\end{figure}

\textcolor[rgb]{0.00,0.00,0.00}{In this section, we will first introduce the proposed \textit{MFRL} framework from an overall perspective. Then we will design the NN structure to solve this EE maximization problem, and propose a meta reinforcement learning scheme for the NN initialization. We also demonstrate the meta-training and meta-adapting algorithms in detail. Finally, we will present the federated learning algorithm and procedures. }

\textcolor[rgb]{0.00,0.00,0.00}{The proposed algorithm starts from the meta-training for initializing the global generalized model at the BS. The initial model is meta-trained using the BS data set. After the initial global model is trained, it will be broadcast to the local UEs for adapting to the new environments. During the meta-adapting, i.e., the fine-tuning process, the local models are trained using a local database, i.e., local CSI, and the local models can be reunited as a global model so that the UEs could learn the knowledge from the experiences of other UEs and improve the global EE. One popular way is to average the distributed models and form a global model, which is called federated learning~\cite{9530714}.  After the local models are averaged by the BS, it would be broadcast to the local UEs which will fine-tune the global model and adapt to the local scenarios. This process will be repeated until the meta-adaptation stage finishes. The overall procedure is shown in Fig.~\ref{FL_framework}}

\subsection{Neural Network Structure Design}
\textcolor[rgb]{0.00,0.00,0.00}{As the aforementioned description, the resource allocation problem can be modeled as a multi-agent markov decision process (MDP), which is mathematically expressed by a tuple, $\langle I, {\cal O}, {\cal A}, {\cal R}, P \rangle$, where $I$ is the number of agents, $N = 1$ degenerates to a single-agent MDP, ${\cal O} $ is the combination set of all observation state, ${\cal A} = {\cal A}_0 \times \dots \times {\cal A}_I$ is the set of actions for each agent, ${\cal R} $ is the reward function, which is related to current observation $O_t = \{o_0, \dots, o_I\} \in {\cal O}$, $A_t = \{a_0, \dots, a_I\} \in {\cal A}$, and $O_{t+1} \in {\cal O}$. Transition probability function is defined as $P :{\cal O} \times {\cal A} \to \mathcal{P}({\cal O})$, with $P(O_{t+1}|O_t,A_t)$ being the probability of transitioning into state $O_{t+1}$ if the environment start in state $O_{t}$ and take joint action $A_t$.}

\textcolor[rgb]{0.00,0.00,0.00}{One of the challenges of using deep reinforcement learning algorithms to solve the problem (\textbf{P0}) is that the resource allocation of the transmit power and subchannel association is the hybrid optimization of the continuous and discrete variables. As the analysis above, the discrete subchannel association parameter can be described by different probabilities to choose different subchannels, thus the discrete variable can be expressed by probability distributions on subchannels, which is generated by a categorical layer. Meanwhile, continuous power optimization is performed by the Gaussian layer, where the mean and variance of the transmit power can be trained. }

\begin{figure}[t]
\centering
\includegraphics[width=\columnwidth]{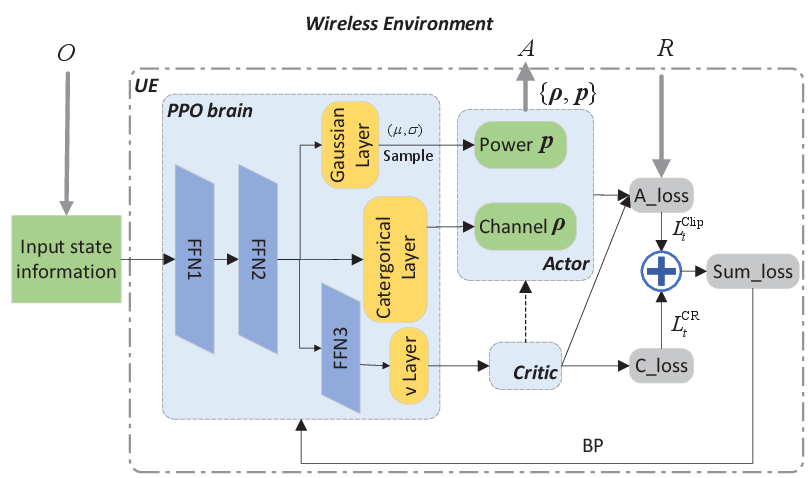}
\centering
\caption{The proposed PPO network structure for the \textit{MFRL} framework.}
\label{network_structure}
\end{figure}

\textcolor[rgb]{0.00,0.00,0.00}{In fact, any deep reinforcement learning algorithms with continuous action space can be applied for training the proposed network structure. Specifically, we apply the PPO algorithm because of its ease of use and robustness, which make it the default algorithm by OpenAI~\cite{PPO_2017}. It is noted that the NN architecture shares parameters between the policy and value function, so that the actor network and critic network share the underlying features in the NN, and simplify the meta learning initialization and model broadcast costs. The corresponding network structure of the local models is illustrated in Fig. \ref{network_structure}.}

In this paper, we define the observation state at training step $t$ for the UEs, which are considered as the agents in the \textit{MFRL} framework,  as $o_{t, i} = \{\{h_{i,n}\}_{\forall n \in N}, t\}$ with dimension~$|o_i|$, where $t$ represents the number of epoch. The variables $t$ can be treated as a low-dimensional \textit{fingerprint} information to contain the policy of other agents~\cite{foerster2017stabilising}, thus enhancing the stationary and the convergence performance of the \textit{MFRL} algorithm.

The action $a_{t, i}$ for the $UE_i$ including the subchannel and the transmit power choice with dimension $|a| = 2$. The Actor network contains a categorical layer with $N$ neurons to determine which subchannel the local UE should access. The continuous transmit power is optimized by a $\boldsymbol \sigma$ layer and a $\boldsymbol \mu$ layer, and the power is sampled according to the probability distribution $N(\boldsymbol \mu,\boldsymbol \sigma^2)$.

Since we aim to maximize the sum EE of the cellular network, here we design a global reward $r_t$, according to the joint action $\boldsymbol{a}_t$ such that encouraging collaboration of UEs. \textcolor[rgb]{0.00,0.00,0.00}{The global reward at training step $t$ can be defined as
\begin{equation}
r_t= 
\begin{cases}
\sum\limits^I_{i=0} r_{i}(t) &\text{if} \sum\limits^I_{i=0}\rho_{i,n} \in \{0, 1\}, \forall i \in I, \forall n \in N;\\
\frac{I^{suc}-I}{I}, &\text{Otherwise},
\end{cases}
\label{global_reward}
\end{equation}
where $I^{suc}$ denotes the number of UEs that satisfies the subchannel assignment constraints, i.e., $\sum^I_{i=0}\rho_{i,n} \in \{0, 1\}, \forall n \in N$. For the assignment that fails to meet the subchannel access requirements, a punishment is set to proportional to the number of failure UEs. Meanwhile, the reward for a successful subchannel assignment is expressed by
\begin{equation}
r_{i}(t)=
\begin{cases}
\xi u_{i,n}(t), &\text{if }\gamma_{i,n} > \gamma_{min};\\
\xi u^{p_{max}}_{i,n}(t), &\text{Otherwise},
\end{cases}
\label{UE_reward}
\end{equation}
where $\xi$ is a constant coefficient, $u^{p_{max}}_{i,n}(t)$ denotes the EE by the maximum transmit power, which means if the UE fails to meet the SNR constraint, it need to use the maximum transmit power to avoid transmission failure. The success rate of $UE_i$ can be defined as $\eta_i = {\beta}_i/{T}$, where ${\beta}_i$ represents the successful resource assignment counts for $UE_i$, and $T$ represents the number of resource allocation counts since the initialization the system.\footnote{\textcolor[rgb]{0.00,0.00,0.00}{Please note that reward is designed as a sum of EE and the punishment, which makes it a dimensionless parameter and we only need to focus on the value of it.}}}

{The objective of the proposed \textit{MFRL} framework is to enable UEs to learn a strategy that maximizes the discount reward, which can be expressed by 
\begin{equation}
R(\tau) = \sum_{t=0}^{\infty} \xi^t r_{t},
\label{eq7}
\end{equation}
where $\tau = (o_0, a_0, ..., o_{T+1})$ is a trajectory, $T$ is the current timestamp, $\xi \in (0, 1)$ represents the discount rate, which denotes the impact of the future reward to the current action.}

\subsection{Policy Gradient in Meta-training}
In the previous work \cite{liang2019spectrum, 9026965, ji2022federated}, the number of UEs in each cluster is fixed, and the training and testing performance are implemented in the same environment. Particularly, the local model is trained by each UE individually for the \textit{MFRL} algorithm, which limits its application, making it hard to adapt to more complicated practical scenarios. The resource allocation model should have the ability to adapt and generalize to different wireless communication environments with different cluster sizes. Hence, the meta reinforcement learning algorithm can be considered to meet the requirement of the generalization.

The meta learning can be implemented in different ways, and in this paper we apply the MAML method for reinforcement learning~\cite{MAML2017}. The meta-training process takes the experience from different tasks, i.e., the resource allocation for different cluster sizes, to initialize a model which can be adopted by UEs in different scenarios and achieve fast adaptation. To take the number of UEs into account, the local observation should include the total number of UEs, i.e., $o_{t, i} = \{\{h_{i,n}\}_{\forall n \in N}, I, t\}$. The task set of resource allocation for UEs is defined as ${\cal T} = \{{\cal T}^{I_k}\}, \forall k \in K$, where $K$ is the number of tasks, $I_k$ is the number of UEs for task $k$. The meta-training process is implemented at the BS, which can use the previous resource allocation experience for different number of UEs to meta-train an initial model. 

At the end of each training epoch, the BS stores the transitions $e^k_{t, i} = \{ (o^k_{t, i}, a^k_{t, i}, r^k_t, o^k_{t+1, i})| i = 0, 1, \dots, I_k-1\}$ acquired from ${\cal T}^{I_k}$ in the central dataset. The transitions $e_{t, i} = (o_{t, i}, a_{t, i}, r_t, o_{t+1, i})$ are sampled from $\cal B$ for calculating the advantage function and the estimated state value function, which are introduced in the following paragraphs. The objective function for training the reinforcement model is to maximize the expected reward for each trajectory as
\begin{equation}
J\left (\pi_\theta\right) =
\mathbb{E}_{\tau\sim \pi_\theta(\tau)}\left [R(\tau)\right]= \int_{\tau} P(\tau|\pi_\theta) R(\tau),
\label{objective_function}
\end{equation}
where $\pi_\theta$ is the parameterized policy, $P(\tau|\pi_\theta) =P (o_0) \prod_{t=0}^{T-1} P(o_{t+1, i} | o_{t,i}, a_{t,i}) \pi_\theta(a_{t,i} | o_{t,i})$ represents the probability of the trajectory $\tau$, $P(o_{t+1, i} | o_{t,i}, a_{t,i})$ is the state transformation probability, $\pi_\theta(a_{t,i} | o_{t,i})$ is the action choice probability, and $P (o_0)$ is the probability of the initial state $o_0$. To optimize the policy, the policy gradient needs to be calculated, i.e., $\theta_{j+1} = \theta_j + \alpha \left. \nabla_{\theta} J(\pi_{\theta}) \right|_{\theta_j}$, where $\alpha$ is the learning rate or the learning step. 

The gradient of the policy can be expressed by a general form as
\begin{equation}
\nabla_{\theta}J(\pi_{\theta}) = \mathbb{E}_{\tau \sim \pi_{\theta}(\tau)}\left[{\sum_{t=0}^{T} \nabla_{\theta} \log \pi_{\theta}(a_{t,i} |o_{t,i}) \Phi_{t,i}}\right],
\label{policy_gradient}
\end{equation}
where $\Phi_{t,i}$ could be denoted as the action-value function $Q^{\pi_\theta}(o,a) = \mathbb{E}_{\tau\sim \pi_\theta(\tau)}\left[R(\tau)| o_0 = o, a_0 = a\right]$, which is the expectation reward for taking action $a$ at state $o$. Although we can use the action-value function to evaluate the action is good or bad, the action-value function $Q^{\pi_\theta}(o,a)$ relies on the state and the action, which means an optimal policy under a bad state may have less action-value than an arbitrary action under a better state. To address this issue, we need to eliminate the influence caused by the state. First, we prove that the state influence elimination will not affect the value of the policy gradient~\cite{GAE_paper}.

\begin{lemma}[Expected Grad-Log-Prob Lemma]

Given $P^{\pi_\theta}$ is a parameterized probability distribution over a random variable $o$, then $\mathbb{E}_{o\sim P^{\pi_\theta}}\left[{\nabla_{\theta} \log P^{\pi_\theta}(o)}\right] = 0.$

\begin{proof} 
For all probability distributions, we have 
\begin{equation}
   \int_o P^{\pi_\theta}(o) = 1. 
\end{equation}
Take the gradient of both side
\begin{equation}
   \nabla_{\theta} \int_o P^{\pi_\theta}(o) = \nabla_{\theta}1 = 0. 
\end{equation}
Thus
\textcolor[rgb]{0.00,0.00,0.00}{\begin{align*}
       &\mathbb{E}_{o\sim P^{\pi_\theta}}\left[{\nabla_{\theta} \log P^{\pi_\theta}(o)}\right]\\
        &= \int_o P^{\pi_\theta}(o) \nabla_{\theta}\log P^{\pi_\theta}(o)\\
        &= \int_o \nabla_{\theta} P^{\pi_\theta}(o)\\
        &= \nabla_{\theta} \int_o P^{\pi_\theta}(o)\\
        &= 0.
\end{align*}}
\label{EGLP}
\end{proof}
\end{lemma}

According to Lemma \ref{EGLP}, we can derive that for any function 
 $b(o_t)$ that only depends on the state, $\mathbb{E}_{a \sim \pi_{\theta}}\left[{\nabla_{\theta} \log \pi_{\theta}(a|o) b(o)}\right]= 0$. Hence, it would cause the same expected value of the policy gradient $\nabla_{\theta}J(\pi_{\theta})$ if we substitute the $b(o)$ into the action-value function $Q^{\pi_\theta}(o,a)$. In fact, we can use the state-value function $V^{\pi_\theta}(o)$ which represents whether the state is good for a higher reward or not. Instead of comparing the action-value function $Q^{\pi_\theta}(o,a)$ of the action $a$ directly, it is more reasonable to substitute the influence of the state into the action-value function. We define the substitution $A^{\pi_\theta}(o,a) = Q^{\pi_\theta}(o,a) - V^{\pi_\theta}(o)$ as the advantage function, which represents whether an action good or bad compared with other actions relative to the current policy. Hence, the value function $\Phi_{t,i}$ can be also denoted as
\begin{equation}
\Phi_{t,i} = Q^{\pi_\theta}(o_{t,i},a_{t,i})-V^{\pi_\theta}(o_{t,i}) = A^{\pi_\theta}(o_{t,i},a_{t,i}).
\label{advantage_function}
\end{equation}

\subsection{Advantage Estimation and Loss Function Design}
Although we express the policy gradient by introducing the advantage function, the challenge is, the action-value function and the state-value function cannot be acquired directly from the experience $e_{t,i}$. Instead, the action-value function can be expressed by the temporal difference form~\cite{sutton2018reinforcement} as $Q^{\pi_\theta}(o_{t,i},a_{t,i}) = r_t + \xi V^{\pi_\theta}(o_{t+1, i})$. In deep reinforcement learning approaches, NNs can be used to estimate the state-value function as $\hat{V}^{\pi_\theta}$, then the estimated advantage function $\hat{A}^{\pi_\theta}(o_{t,i},a_{t,i}) = \delta^V_{t,i} = r_t + \xi \hat{V}^{\pi_\theta}(o_{t+1,i}) -\hat{V}^{\pi_\theta}(o_{t,i})$ can be derived. However, the bias for this estimation is high, which restricts the training and convergence performance. To overcome this issue, generalized advantage estimation (GAE)~\cite{GAE_paper} can be applied to estimate the advantage function for multi-steps and strike a tradeoff between the bias and variance. The GAE advantage function is denoted by
\begin{equation}
    A^{\text{GAE}}(o_{t,i},a_{t,i}) =  \sum \limits _{l=0}^{T-t}(\lambda\xi)^l\delta^V_{t+l,i},
\label{GAE_function}
\end{equation}
where $\lambda \in (0,1]$ is the discount factor for reducing the variance of the future advantage estimation.

The actor network is optimized by maximising  $L_{AC} = \mathbb{E}_{\tau \sim \pi_{\theta}(\tau)}\left[ratio_{t,i} \times A^{\text{GAE}}(o_{t,i},a_{t,i})\right]$, where $ratio_{t,i} =\frac{\pi_{\theta}(a_{t,i} |o_{t,i})}{\pi_{\theta_{\text{old}}}(a_{t,i} |o_{t,i})}$ is the action step. However, too large action step could lead to an excessively large policy update, hence we can clip this step and restrict it. The clipped actor objective function is expressed by
\begin{equation}
    L^{\text{Clip}}_t = \min\left(
ratio_{t,i}  \times A^{\text{GAE}}(o_{t,i},a_{t,i}),\;
g(\epsilon, A^{\text{GAE}}(o_{t,i},a_{t,i}))\right),
\label{clipped_loss}
\end{equation}
where 
\begin{equation}
g(\epsilon, A) =
\begin{cases}
(1 + \epsilon) A, &A \geq 0;\\
(1 - \epsilon) A & A < 0,
\end{cases}
\label{clip}
\end{equation}
in which the $\epsilon$ is a constant value representing the clip range. The clip operation have been proved to improve the robustness~\cite{PPO_2017}.  

The loss $L_{\text{CR}}$ for the critic network is to minimize the gap between the estimated state-value function and discount sum reward, which can be expressed by
\begin{equation}
    L^{\text{CR}}_t =\left\Vert r_t + \hat{V}^{\pi_\theta}(o_{t+1,i}) - \hat{V}^{\pi_\theta}(o_{t,i}) \right\Vert^ 2.
\label{TD}
\end{equation}

Combining the objective of the actor network and critic network, we can express the overall objective as
\begin{equation}
L=\arg \min_{\theta} \mathbb{E}_{t}\left[L^{\text{Clip}}_t-c_1L^{\text{CR}}_t+c_2E_t\right],
\label{overall_loss}
\end{equation}
where $E_t$ represents an entropy bonus to ensure sufficient exploration, $\theta$ is the weights for the PPO network, $c_1$ and $c_2$ are weight parameters for the estimation of value function and entropy, respectively. Then the initial model will be updated by the stochastic gradient (SG) ascent approach. The details of the meta-training algorithm is shown in \textbf{Algorithm \ref{algo1}}.

\begin{algorithm}[t]
\caption{Meta-training algorithm.}
\label{algo1}
\begin{algorithmic}[1]
\STATE \textbf{Input}: The task set ${\cal T} = \{{\cal T}^{I_k}\}, \forall k \in K$, BS memory ${\cal M}$, BS batch ${\cal B}$;\
\STATE Initialize the PPO network $\theta$;\
\FOR{each epoch $t$}
\FOR{each meta task $k$}
\STATE The BS acquire the experience $e^k_{t, i} = \{ (o^k_{t, i}, a^k_{t, i}, r^k_t, o^k_{t+1, i})| i = 0, 1, \dots, I_k-1\}$ from all UEs and store the transitions in central dataset ${\cal M}$;\
\ENDFOR
\STATE Sample the transitions in the BS batch ${\cal B}$;\
\STATE Update the global PPO network by SG ascent with Adam:
$\theta \leftarrow \theta + \alpha_{\text{meta}}\nabla_{\theta}L$;\
\ENDFOR
\STATE \textbf{Return}: Pre-trained global model $\theta$.
\end{algorithmic}
\end{algorithm}
\subsection{\textcolor[rgb]{0.00,0.00,0.00}{Meta-Adapting Process}}
Unlike the meta-training process where the BS stores the transitions and uses these experiences to train a global model, the local UE can train its own model based on its own observations and experience during the meta-adaptation process. Compared with supervised learning which requires sufficient data set and pre-knowledge of the system, the proposed \textit{MFRL} framework can train the local model with the local CSI data which is required by interacting with the environment, thus not only offloading the computational pressure to the UEs, but also lower the transmission overhead significantly.

As the local models are inherited from the global model, the network structure, the observation state space, the action, and the reward are defined the same as Section~\uppercase\expandafter{\romannumeral3}. Considering that the $i$-th UE interacts with the environment at adapting epoch $j$, i.e., observes the state $o_{j,i}$, and takes action according to current policy $\pi(\theta_{j,i})$. Then the $i$-th UE receives the reward $r_j$ and observes the next state $o_{j+1,i}$. The transition $e_{j,i} = (o_{j,i}, a_{j,i}, r_j, o_{j+1,i})$ is stored in its local memory ${\cal M}_i$ which can be sampled in the batch to train the local models. The advantage is estimated using the GAE method and the loss function is the same as the meta-training process. The details of the meta-adapting process are described in \textbf{Algorithm \ref{algo2}}.

\begin{algorithm}[t]
\caption{Meta-adapting algorithm.}
\label{algo2}
\begin{algorithmic}[1]
\STATE \textbf{Input}: The pre-trained global model $\theta$, number of UEs $I$, local memory ${\cal M}_i$ and batch ${\cal B}_i$ for each UE;\
\STATE Initialize the local models $\theta_{0,i} \leftarrow \theta, \forall i \in I$;\
\FOR{each epoch $j$}
\FOR{each D2D pair $i$}
\STATE Collect set of trajectories ${\cal M}_i$ by running policy $\pi_{j,i} = \pi(\theta_{j,i})$ in the environment;\
\STATE Compute advantage estimations ${A}^{\text{GAE}}(o_{j,i},a_{j,i})$ based on current state-value function $\hat{V}^{\pi_\theta}(o)$ and reward $r_j$;\
\STATE Update the PPO network by maximizing the objective function:\\
$\theta_{j+1, i} = \mathop{\arg\max}\limits_{\theta_i} {\frac{1}{T}\sum\limits^T_{j=0}\left(L^{\text{Clip}}_j-c_1L^{\text{CR}}_j+c_2E_j\right)}$;\
\ENDFOR
\ENDFOR
\end{algorithmic}
\end{algorithm}

\subsection{Global Averaging of Local Models}
Unlike the meta-training process that the BS uses the centralized replay memory that collects from all UEs to train the global model, the local UEs can only access their local memories during the meta-adaptation process, which affects the robustness of the local models when encountering unfamiliar scenarios. To enable the individual models at each UE can be benefited from other UEs, the federated learning technique can be applied. 

The local model is averaged to a global model, then the global model is broadcast to UEs and the UEs will continue to train the new global model locally. By averaging the models, each UE is able to benefit from the experience of other UEs, since the weights direct correspond to the experience and memory. Mathematically, the model averaging process at the central BS can be denoted as
\begin{equation}
\overline{\boldsymbol W} = \frac{\sum^I_{i=1} |{\cal B}_i| \boldsymbol W_i}{\sum^I_{i=1} |{\cal B}_i| },
\label{FL}
\end{equation}
where $|{\cal B}_i|$ represents the number of number of elements in ${\cal B}_i$. The average algorithm shows that the averaged model will learn more from the model with more training cases. 
However, in the proposed \textit{MFRL} framework, we assume that UEs share the team stage reward, which means the replay memory of each UE has an equivalent size. To ensure that the averaged model can benefit from the model that caters to the needs of QoS, we further revised the averaging algorithm that considers the success rate, which is denoted by
\begin{equation}
\overline{\hat{\boldsymbol W}} = \frac{\sum^I_{i=1}\eta_i \boldsymbol W_i}{\sum^I_{i=1}\eta_i},
\label{eq12}
\end{equation}
where $\eta_i$ is the resource allocation success rate for $UE_i$ as defined in Section~\uppercase\expandafter{\romannumeral2}.

\section{Numerical Results}
\begin{table}[t]
\begin{center}
\caption{Environment Parameters}
\begin{tabular}{|c|c|}
\hline
\textbf{Parameter}&
\textbf{Value} \\
\hline
Antenna gain of the BS &
8dB\\
\hline
Antenna gain of the UEs &
3dB\\
\hline
Noise figure at the BS &
5dB\\
\hline
Noise figure at the UEs &
9dB\\
\hline
Number of UEs $I$&
6\\
\hline
Number of UEs for different tasks in meta learning&
[2, 4, 8]\\
\hline
Number of subchannels $N$&
10\\
\hline
Height of antenna of the UEs &
1.5\meter\\
\hline
Number of subcarriers in a RB $K$&
12\\
\hline
Carrier frequency $f_n, \forall n \in N$&
6G\hertz\\
\hline
Cellular transmit power range&
$[0, 24]$dBm\\
\hline
Minimum SINR requirements for BS $\gamma^C_{min}$&
5 dB\\
\hline
Noise power spectral density of indoor scenario &
-160 dBm/\hertz\\
\hline
Noise power spectral density of urban micro scenario &
-170 dBm/\hertz\\
\hline
Noise power spectral density of urban macro scenario &
-180 dBm/\hertz\\
\hline
Noise power spectral density of rural macro scenario &
-185 dBm/\hertz\\
\hline
Shadowing distribution&
Log-normal\\
\hline
Pathloss and shadowing update&
Every 100m\second\\
\hline
Fast fading update&
Every 1m\second\\
\hline
\end{tabular}
\end{center}
\label{tab1}
\end{table}

\begin{figure}[t]
\centering
\includegraphics[width=\columnwidth]{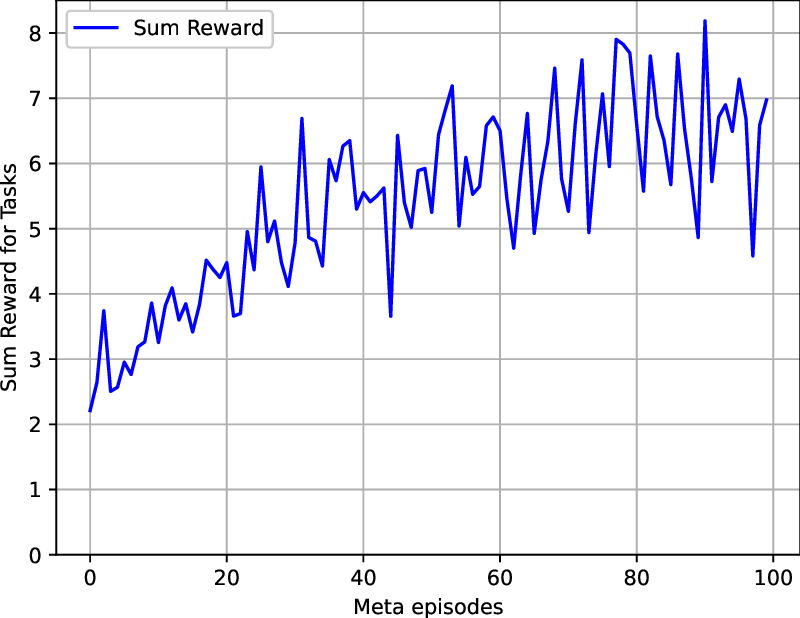}
\caption{meta-training reward over the meta-training episodes. The curve represents the sum reward the agent gets from different tasks.}
\label{Meta_reward}
\end{figure}

\begin{figure*}[t]
\subfigure[Indoor scenario.]{
\begin{minipage}[t]{0.33\linewidth}
\includegraphics[width=0.95\columnwidth]{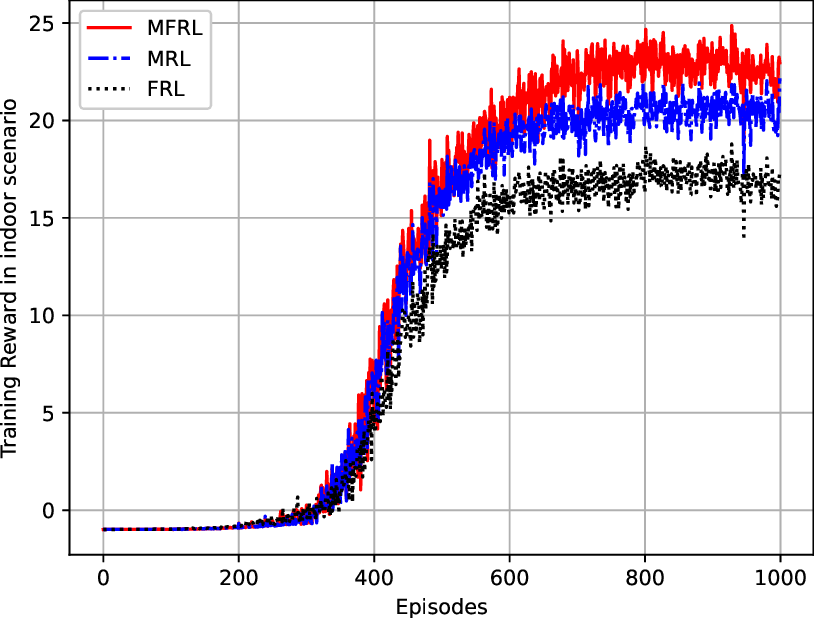}
\end{minipage}%
}%
\subfigure[Urban macro scenario.]{
\centering
\begin{minipage}[t]{0.33\linewidth}
\includegraphics[width=0.95\columnwidth]{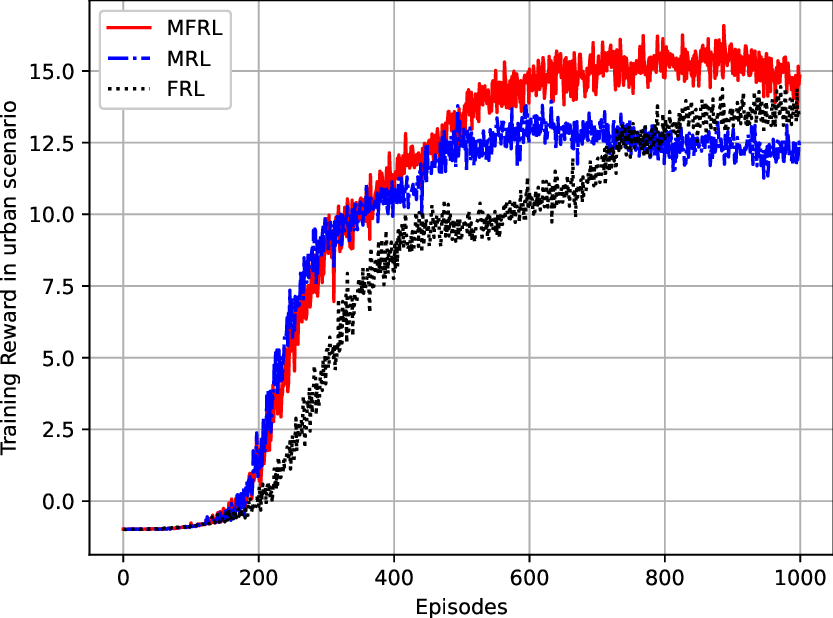}
\end{minipage}%
}%
\subfigure[Rural macro scenario.]{
\centering
\begin{minipage}[t]{0.33\linewidth}
\includegraphics[width=0.95\columnwidth]{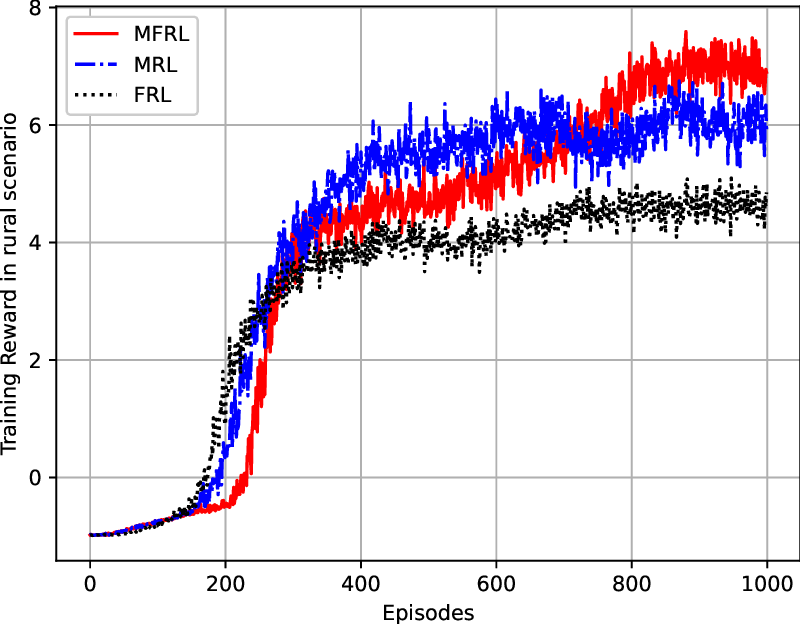}
\end{minipage}%
}%
\caption{Training performance comparison of the proposed algorithm and benchmarks in three different scenarios.}
\label{training_performance}
\end{figure*}

\begin{figure*}[ht]
\subfigure[Indoor scenario.]{
\begin{minipage}[t]{0.33\linewidth}
\includegraphics[width=0.95\columnwidth]{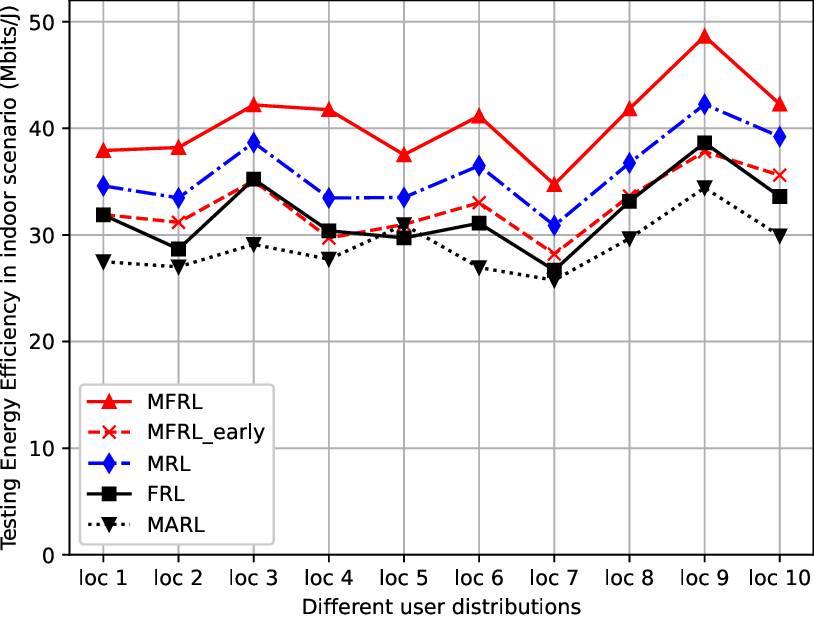}
\end{minipage}%
}%
\subfigure[Urban macro scenario.]{
\centering
\begin{minipage}[t]{0.33\linewidth}
\includegraphics[width=0.95\columnwidth]{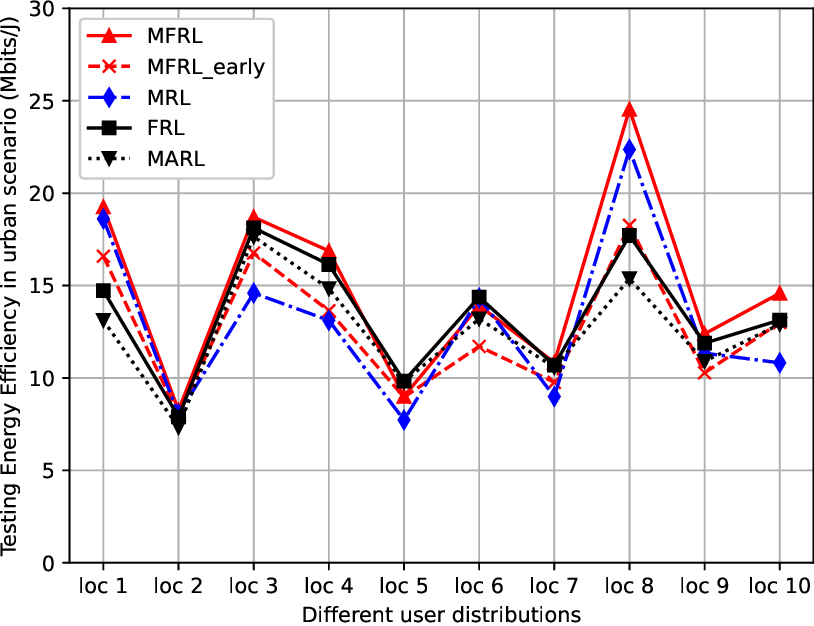}
\end{minipage}%
}%
\subfigure[Rural macro scenario.]{
\centering
\begin{minipage}[t]{0.33\linewidth}
\includegraphics[width=0.95\columnwidth]{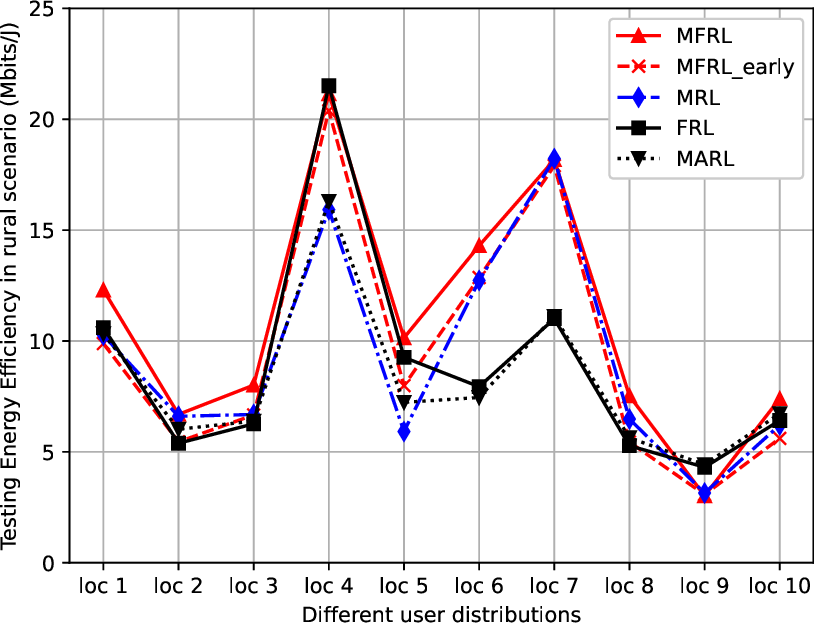}
\end{minipage}%
}%
\caption{Testing snapshots of the proposed algorithm and benchmarks in three different scenarios.}
\label{test_EE}
\end{figure*}

We consider a communication scenario underlying a single cellular network. For the meta-training process, we adopt the urban micro (street canyon) scenario in~\cite{3gpp_38.901}. For the meta-adaptation process, the pre-trained models are trained and fine-tuned in the indoor scenario, the urban macro scenario, and the rural macro scenario. For all of the scenarios, the BS is fixed at the center of the considered square. We also adopt the simulation assumptions in~\cite{3gpp_38.901} to model the channels. To enable the mobility of UEs, we assume that the UEs can move with the speed from 0 meters per second (m/s) to 1 m/s within the square. \textcolor[rgb]{0.00,0.00,0.00}{Each subcarrier has $\Delta f = 2^\psi \cdot 15$ kHz spacing, where $\psi$ denotes an integer. A resource block usually consists of 12 consecutive subcarriers~\cite{3gpp_38.211}, hence we set the bandwidth set of the subchannels as [0.18, 0.18, 0.36, 0.36, 0.36, 0.72, 0.72, 0.72, 1.44, 1.44] M$\hertz$.} The rest of the parameters of the proposed simulation environment are listed in Table~\ref{tab1}. 

The network structure of local models is shown in Fig.~\ref{network_structure}. The state information is fed in two fully connected feed-forward hidden layers, which contain 512 and 256 neurons respectively. Then the PPO network diverges to actor networks and critic networks. The actor branch contains two layers for channel choice and power optimization independently, while the critic branch includes an additional hidden layer with 128 neurons, following which is the value layer for estimating the advantage function for the output of the actor network. The meta-training rate for different number of users is $5\mathrm{e}^{-7}$, while the learning rate for meta adaptation is $1\mathrm{e}^{-6}$. The meta learning rate is set relatively small to avoid the overfitting of the meta model for some specific tasks. The weight for the loss of the value function $c_1$ and entropy $c_2$ are set as 0.5 and 0.01, respectively. The sample batch size is 256, and the discount rate for the future reward $\xi$ is set to $0.9$. The discount factor for the advantage function $\lambda = 0.98$ in Eq. (11) is set according to~\cite{PPO_2017}.

To verify the performance of the proposed \textit{MFRL} framework with the following benchmarks:
\begin{enumerate}[1)]
\item \textbf{MRL}: Meta reinforcement learning benchmark. The local models are pre-trained and inherited from the global model, but the local models are not averaged by federated learning.
\item \textbf{FRL}: Federated reinforcement leanring benchmark. The local models are trained from the random initialization and averaged by the federated learning every 100 episodes.
\item \textbf{MFRL\_early}: The early model of the proposed \textit{MFRL} framework. The models are stored at half of the meta-adaptation period, i.e., store the local models at 500 episodes to evaluate the fast-adaptation performance of the proposed framework at the early stage.
\item \textbf{MARL}: The multi-agent reinforcement learning benchmark~\cite{liang2019spectrum}. The local models are trained from random initialization and are not averaged by the federated learning technique. Each UE learns the policy according to the local observations and receives the global reward, but cannot communicate the model with the centralized cloud or other UEs.
\end{enumerate}

Fig. \ref{Meta_reward} demonstrates the reward for different tasks (with different amounts of users) during the meta-training process. Particularly, the meta reward is the sum of the reward of the resource allocation tasks for 2, 4, and 8 UEs in the urban micro scenario. The increase in the meta reward demonstrates the effectiveness of the meta-training. It is also noted that with the meta-training step increasing over 100 episodes, the sum reward keeps stable. This is because the meta-training process is to train a global and generalized model which can be adapted to different tasks, but the performance of the generalized model itself cannot be as well as the models for the specific tasks.

Fig. \ref{training_performance} shows the training reward comparison over different episodes of meta-training, from which we can see that the meta-training could lead to faster convergence and higher rewards. Due to the punishment, the reward for all schemes is low at the beginning of the training period. With the execution of the training progress, the proposed algorithms with meta learning can achieve faster convergence and higher training reward, while the conventional benchmark needs more iterations to find the appropriate actions to converge. \textcolor[rgb]{0.00,0.00,0.00}{The improved training performance verifies the fast adaptation by the meta learning is robust to different scenarios. }

\begin{figure}[t]
\centering
\includegraphics[width=\columnwidth]{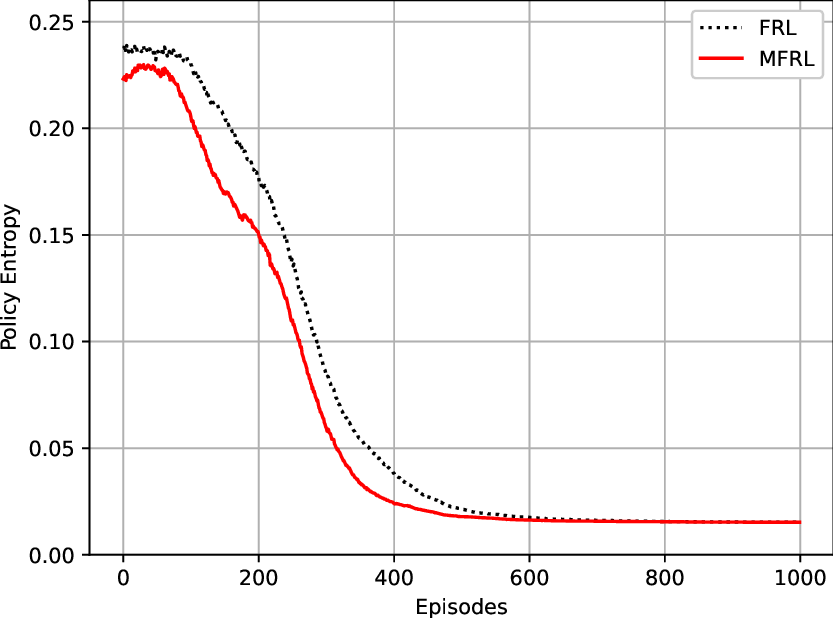}
\centering
\caption{Policy entropy of the \textit{MFRL} and FRL schemes in the indoor scenario.}
\label{entropy}
\end{figure}

\begin{figure}[t]
\centering
\includegraphics[width=\columnwidth]{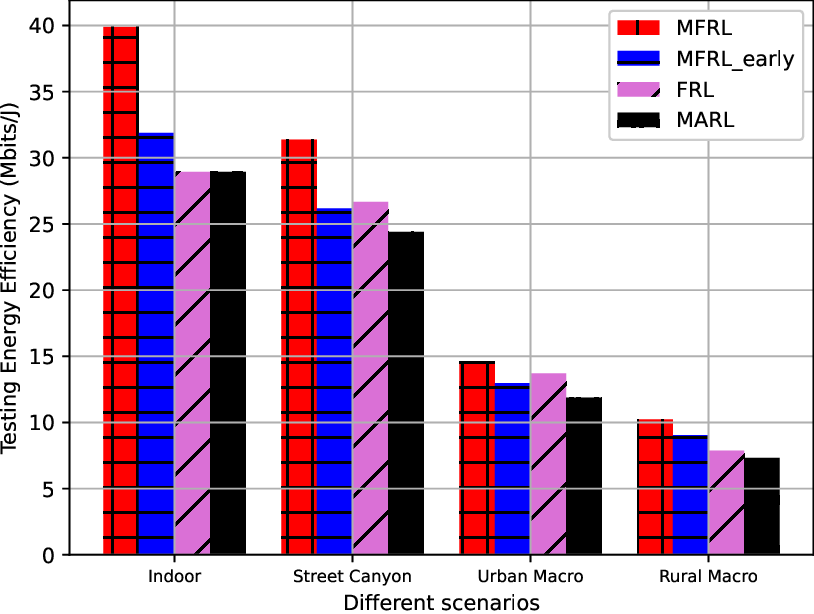}
\centering
\caption{Testing averaged EE performance of 100 random user distributions over the number of model averaging times.}
\label{scenarios}
\end{figure}

\begin{figure}[t]
\centering
\includegraphics[width=\columnwidth]{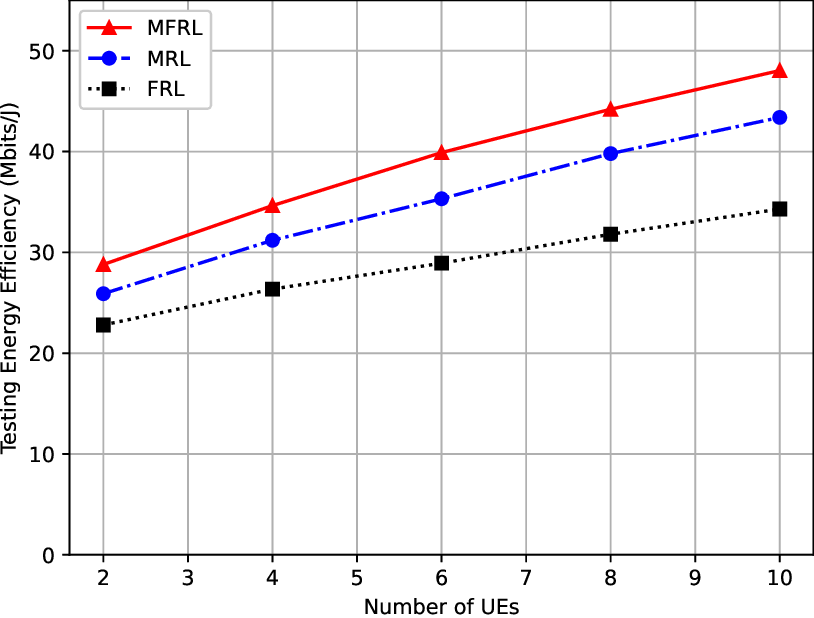}
\centering
\caption{{Testing energy efficiency over the different number of users.}}
\label{User_number}
\end{figure}

\textcolor[rgb]{0.00,0.00,0.00}{To further verify the robustness of the trained local models, we set different simulation settings under each scenario. At each random testing user distribution, the system EE is averaged by 100 testing steps with fast-fading channel updates.} Fig. \ref{test_EE} illustrates the testing performance for 10 random user distributions. The proposed algorithm outperforms other reinforcement learning benchmarks in terms of average system EE. We also store the local models at 500 episodes to test the performance of the algorithms at the early training stage. As expected, the proposed \textit{MFRL} framework outperforms the MRL and FRL algorithms. \textcolor[rgb]{0.00,0.00,0.00}{Moreover, even if MFRL\_early models are only trained half of the whole training period, they still provide good performances compared with the models that are not pre-trained, which verifies the fast adaptation ascendancy of the meta learning.}

To evaluate the convergence speed and the stability of the policy, and verify the fast adaptation performance of the proposed \textit{MFRL} framework, we use the policy entropy as the measure. The policy entropy is an \textcolor[rgb]{0.00,0.00,0.00}{dimensionless} index in policy gradient based reinforcement learning algorithms, to measure the randomness of a policy. As shown in Fig.~\ref{entropy}, the lower entropy of the \textit{MFRL} algorithm verifies that meta learning can speed up the training process and achieve convergence earlier. The \textit{MFRL} framework also achieves a similar lower entropy and faster convergence compared with the benchmarks in other scenarios, and the results are omitted due to space limitations.

Fig. \ref{scenarios} concludes the sum EE in different scenarios. The results are averaged according to 100 random user distributions. It is clear that the proposed \textit{MFRL} framework achieves the highest sum EE in all of the scenarios, which verifies the robustness of the proposed scheme. Additionally, although the models for the MFRL\_early benchmarks are trained half of the whole adapting period, it still achieves better performance compared with the FRL and MARL models. The \textit{MFRL} framework and the FRL scheme enable the UEs to corporate with each others and benefit the local models, hence also improving the overall system EE. 

Fig. \ref{User_number} shows the testing sum EE of the system over a different number of users. Note that for different users, the training parameters may differ slightly for the best performance. It is obvious that as the number of UEs increases, more subchannels can be accessed and the sum system EE can be improved. However, the improvement slows down as the number of UEs increases, since the bandwidth of subchannels in the proposed scenario is not equal, and when the number of UEs is less than the subchannels, it would access the subchannel with larger bandwidth for higher EE.

\section{Conclusion}
In this paper, a distributed energy-efficient resource allocation scheme was developed. The system energy efficiency was maximized by jointly optimizing the channel assignment and the transmit power of user equipments. The formulated non-convex problem was solved by the proposed robust meta federated reinforcement learning framework to overcome the challenge of the computational complexity at the base station and the transmission cost by the local data. \textcolor[rgb]{0.00,0.00,0.00}{Quantity analysis and numerical results showed that the meta training model has good generalization ability under different scenarios, even if the scenarios and tasks are different. Meanwhile, the combination of federated learning and meta learning with reinforcement learning enables the decentralized algorithm a better performance on convergence and robustness.}


\bibliography{Reference}

\end{document}